\documentclass[letterpaper]{article} % DO NOT CHANGE THIS
\usepackage{aaai25}  % DO NOT CHANGE THIS
\usepackage{times}  % DO NOT CHANGE THIS
\usepackage{helvet}  % DO NOT CHANGE THIS
\usepackage{courier}  % DO NOT CHANGE THIS
\usepackage[hyphens]{url}  % DO NOT CHANGE THIS
\usepackage{graphicx} % DO NOT CHANGE THIS
\usepackage{natbib}  % DO NOT CHANGE THIS AND DO NOT ADD ANY OPTIONS TO IT
\usepackage{caption} % DO NOT CHANGE THIS AND DO NOT ADD ANY OPTIONS TO IT
\usepackage{amsmath}
\usepackage{amssymb}
\usepackage{booktabs}
\usepackage{multirow}
\usepackage{hhline}
\usepackage{dsfont}
\usepackage[table,xcdraw,dvipsnames]{xcolor}
\usepackage{hhline}
\usepackage{makecell}
\usepackage{soul}
\usepackage{xcolor, pifont}
\usepackage{rotating}

\usepackage{algorithm}
\usepackage{algorithmic}

\usepackage{newfloat}
\usepackage{listings}
\DeclareCaptionStyle{ruled}{labelfont=normalfont,labelsep=colon,strut=off} % DO NOT CHANGE THIS
\floatstyle{ruled}
\newfloat{listing}{tb}{lst}{}
\floatname{listing}{Listing}
\title{Rank-O-ToM: Unlocking Emotional Nuance Ranking\\to Enhance Affective Theory-of-Mind}
\author{
    JiHyun Kim\textsuperscript{\rm 1}\equalcontrib,
    JuneHyoung Kwon\textsuperscript{\rm 1}\equalcontrib,
    MiHyeon Kim\textsuperscript{\rm 1}\equalcontrib,
    Eunju Lee\textsuperscript{\rm 1}\equalcontrib,
    YoungBin Kim\textsuperscript{\rm 1}\thanks{Corresponding author}
}
\affiliations{
    \textsuperscript{\rm 1}Chung-Ang University  
    \{rlawlgus251, dirchdmltnv, mh10967, dmswn5829, ybkim85\}@cau.ac.kr
}

\usepackage{bibentry}
\begin{document}

\maketitle

\begin{abstract}
\vspace*{-2mm}
Facial Expression Recognition (FER) plays a foundational role in enabling AI systems to interpret emotional nuances, a critical aspect of affective Theory of Mind (ToM). However, existing models often struggle with poor calibration and limited capacity to capture emotional intensity and complexity. To address this, we propose \textit{\textbf{Ranking the Emotional Nuance for Theory of Mind (Rank-O-ToM)}}, a framework that leverages ordinal ranking to align confidence levels with emotional spectra. By incorporating synthetic samples reflecting diverse affective complexities, Rank-O-ToM enhances the nuanced understanding of emotions, advancing AI’s ability to reason about affective states.
\vspace*{-1mm}
\end{abstract}

\vspace*{-4mm}
\section{Introduction}
% \vspace*{-1mm}
Bridging the gap between AI and human understanding requires accurate emotional recognition. To foster trust and empathetic communication, AI models must recognize basic emotions and interpret nuances—a capability rooted in the affective Theory of Mind (ToM)~\cite{stewart2016theory, poletti2012cognitive}. Facial Expression Recognition (FER) is crucial, as facial expressions universally convey emotions and intentions~\cite{baron1997another}. By interpreting these expressions, AI models can respond effectively to emotional cues in applications like compassionate healthcare~\cite{tehranineshat2019compassionate} and adaptive education~\cite{lin2024artificial}.

\begin{figure}[t!]
  \centering
  \includegraphics[width=0.95\columnwidth]{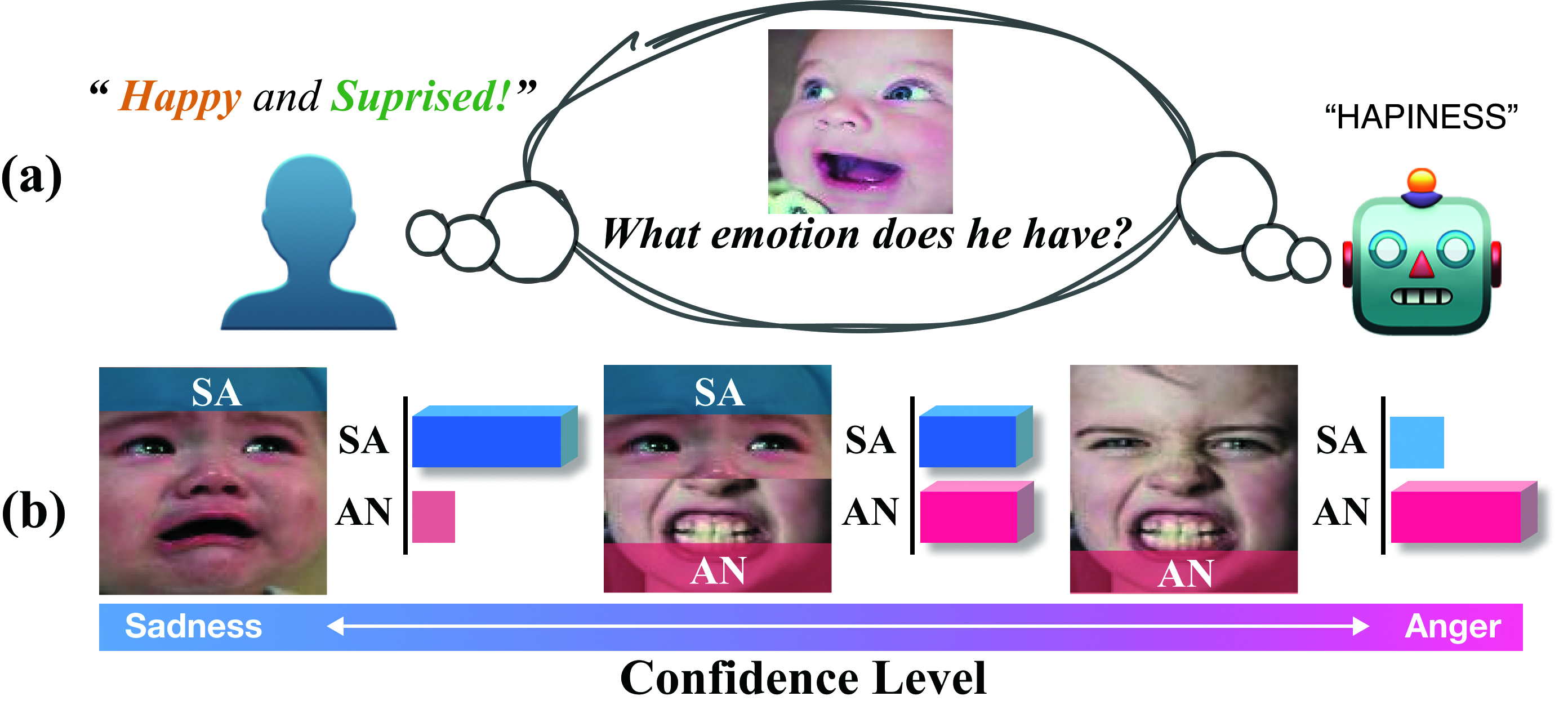}
  \vspace*{-2mm}
  \caption{(a) Affective ToM challenge: interpreting nuanced emotional states, such as blended emotions. (b) Rank-O-ToM blends basic expressions into synthetic samples with ranked confidence scores to capture the emotional spectrum.}
  \vspace*{-3mm}
  \label{fig:motivation}
\end{figure}

At the heart of affective ToM is the ability to understand emotions along with their intensity and complexity~\cite{gabriel2021cognitive}, as shown in Figure~\ref{fig:motivation} (a). Humans often convey subtle and blended emotions, such as simultaneous happiness and surprise, with varying intensity~\cite{du2014compound}, which helps infer deeper mental states. However, existing FER frameworks rely heavily on datasets with single basic emotion labels (e.g., happiness, anger)~\cite{li2017reliable}, limiting their ability to generalize and interpret emotional intensity and complexity. An advanced framework that can recognize nuanced emotional states and produce appropriate confidence scores (i.e., the model’s certainty about its predictions for basic emotion categories) while reflecting varying intensity and complexity is urgently needed in the field of affective ToM.

Inspired by human cognition to interpret emotional nuances, we propose a novel FER framework, \textbf{\textit{Ranking the Emotional Nuance for Theory of Mind (Rank-O-ToM)}}, addressing affective granularity. Our method synthesizes diverse training samples by blending basic emotions to capture real-world affective complexity. To interpret these variations, the model employs a ranking mechanism aligning confidence levels with emotional intensity and clarity, enabling it to distinguish subtle and blended affective cues. Grounded in these principles, Rank-O-ToM enhances AI's capacity for nuanced affective reasoning, advancing affective ToM.

\begin{table*}[t!]
\centering
\resizebox{\linewidth}{!}{%
% \large{
\setlength{\tabcolsep}{5.5pt}
\begin{tabular}{c|cccc|cccc|cccc}
% \toprule
% \Xhline{2\arrayrulewidth}
\hline
                         & \multicolumn{4}{c|}{\textbf{RAF-DB}}                                                       & \multicolumn{4}{c|}{\textbf{FERPlus}}                                                       & \multicolumn{4}{c}{\textbf{AffectNet}}                                                     \\ \hhline{~|------------|}
\multirow{-2}{*}{Method} & \textit{Acc} ↑ & \textit{AECE} ↓ & \textit{MCE} ↓  & \cellcolor[HTML]{EFEFEF}\textit{ECE} ↓  & \textit{Acc} ↑ & \textit{AECE} ↓ & \textit{MCE} ↓   & \cellcolor[HTML]{EFEFEF}\textit{ECE} ↓  & \textit{Acc} ↑ & \textit{AECE} ↓ & \textit{MCE} ↓  & \cellcolor[HTML]{EFEFEF}\textit{ECE} ↓  \\ %\hhline{|-|------------|}
\hline %\hline
SCN~\shortcite{wang2020suppressing} & 88.72              & \underline{5.51}    & \underline{18.18}   & \cellcolor[HTML]{EFEFEF}5.46          & 79.32              & 15.79         & 34.94          & \cellcolor[HTML]{EFEFEF}15.84         & 48.31              & 21.20         & 13.56         & \cellcolor[HTML]{EFEFEF}21.21         \\
EAC~\shortcite{zhang2022learn}   & 89.96              & 6.23          & 34.48         & \cellcolor[HTML]{EFEFEF}\underline{4.87}    & 83.99              & \underline{12.64}   & 36.16          & \cellcolor[HTML]{EFEFEF}\underline{12.71}   & {52.26}        & 25.91         & \textbf{0.98} & \cellcolor[HTML]{EFEFEF}25.92         \\
RAC~\shortcite{zhang2024leave} & \underline{92.11}        & 25.4          & 27.4          & \cellcolor[HTML]{EFEFEF}25.39         & {\underline{85.92}}     & 19.16         & \underline{24.50}     & \cellcolor[HTML]{EFEFEF}19.73         & \underline{54.14}     & \textbf{9.03}    & 11.39         & \cellcolor[HTML]{EFEFEF}\underline{9.20}    \\
\textbf{Ours}       & \textbf{95.00}     & \textbf{3.06} & \textbf{3.19} & \cellcolor[HTML]{EFEFEF}\textbf{2.74} & {\textbf{86.44}}        & {\textbf{9.57}} & {\textbf{18.08}} & \cellcolor[HTML]{EFEFEF}{\textbf{9.59}} & {\textbf{54.84}}              & {\underline{10.28}} & {\underline{10.47}}    & \cellcolor[HTML]{EFEFEF}{\textbf{5.93}} \\ 
\hline
\end{tabular}
}
\vspace*{-2mm}
\caption{Comparison of Top-1 accuracy (Acc, \%), AECE (\%), MCE (\%), and ECE (\%) for FER methods on RAF-DB, FERPlus, and AffectNet. Top performances are bolded, and second-best results are underlined.}
\label{tab1}
\end{table*}

\vspace*{-2mm}
\section{Method}
% \vspace*{-1mm}
AI models struggle to interpret complex affective states due to the limited diversity of basic emotions in FER datasets. To address this, we propose Rank-O-ToM, combining synthetic emotion blending and a ranking-based loss function, as shown in Figure~\ref{fig:motivation} (b), to enhance emotional granularity.

\noindent \textbf{Synthetic Sample Generation.} Humans express emotions through subtle variations in facial regions (e.g., upper face for surprise, lower face for happiness)\cite{lien2000detection}. To capture this, we synthesize samples by horizontally blending images annotated with basic emotions, enriching the training data to reflect real-world affective diversity better. Additionally, we incorporate samples from face recognition (FR) datasets to further enhance diversity (details in Appendix).

\begin{figure}[t]
\vspace*{-5mm}
  \centering
  \includegraphics[width=0.95\columnwidth]{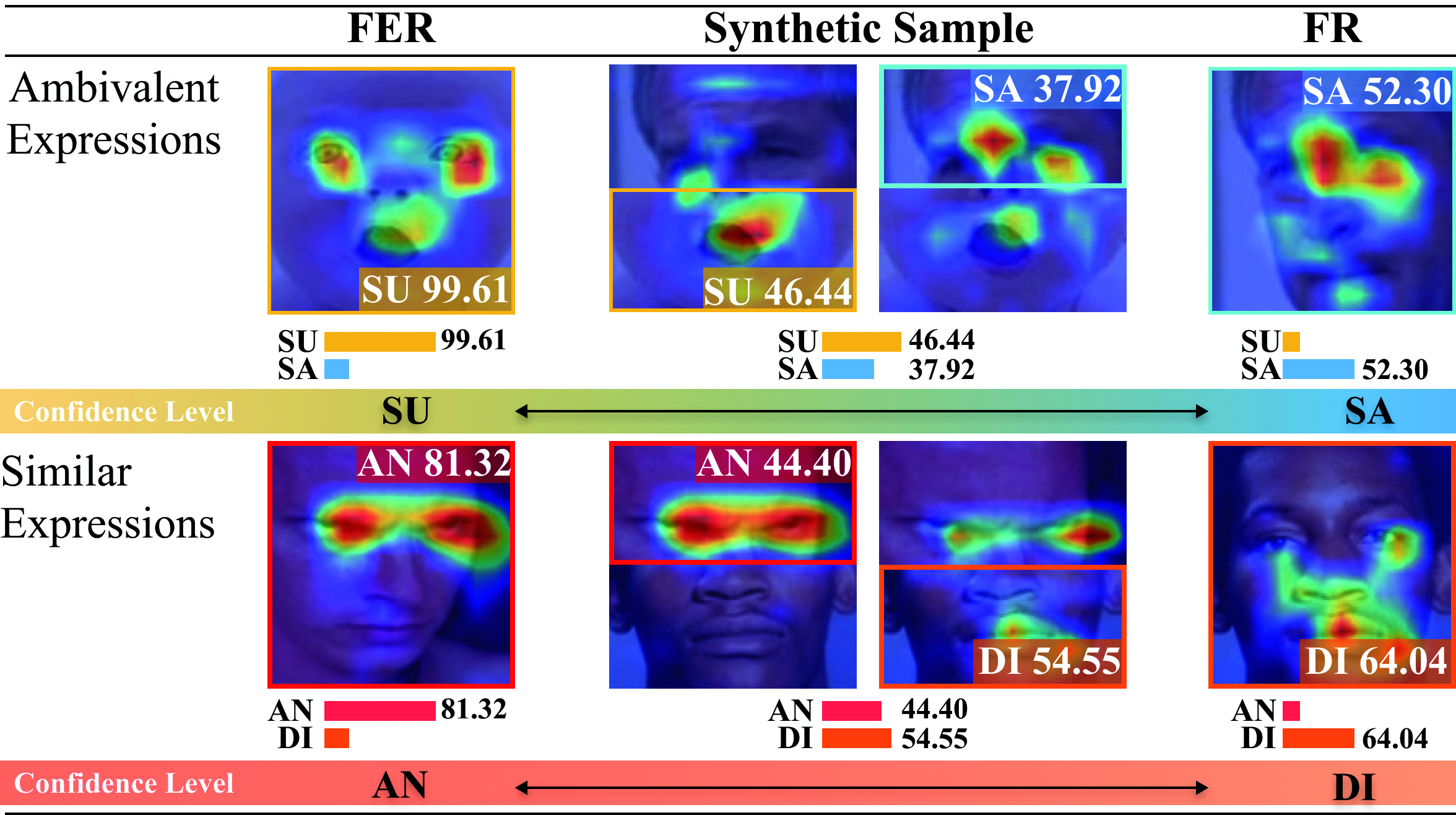}
  \vspace*{-2mm}
  \caption{CAMs and confidence scores for FR, FER, and synthetic samples, showing activations for ambivalent (top) and similar (bottom) expressions with predicted labels.}
  \vspace*{-3.5mm}
  \label{fig:CAM}
\end{figure}

\noindent \textbf{Ranking Loss for Ordinal Relationships.} We propose a ranking-based loss function aligning the model’s confidence levels with the hierarchical structure of human perception for ordinal relationships between affective states \cite{barrett2009affect, cittadini2023affective}. This ensures the model assigns higher confidence scores to original samples (representing clearer or more intense emotions) and lower scores to synthetic samples (representing blended or less intense emotions). The loss function is as follows:
\begin{align}
    \mathcal{L}_{rank} &= max(0, \max_{c_1} p^{\mathrm{syn}}_{c_1} - \max_{c_1} p^{\mathrm
{fer}}_{c_1} + \delta) \notag\\ & \quad + max(0, \max_{c_2} p^{\mathrm{syn}}_{c_2} - \max_{c_2} p^{\mathrm{fr}}_{c_2} + \delta)
    \label{eq:9}
\end{align}

\vspace*{-1.5mm}

Here, $p^{\mathrm{syn}}_{c1}$ and $p^{\mathrm{fer}}_{c1}$ are the confidence scores for synthetic and original samples in the basic emotion category $c_1$, respectively, and $p^{\mathrm{syn}}_{c2}$ and $p^{\mathrm{fer}}_{c2}$ are their counterparts for category $c_2$. By enforcing a meaningful separation with the margin $\delta$, the loss ensures the model assigns higher confidence to clear, intense emotions in original samples while producing lower confidence for blended or less intense emotions in synthetic samples. This mechanism enables the model to capture subtle variations in affective states and maintain consistent confidence across diverse emotional expressions.

\vspace*{-1.5mm}
\section{Experiment} 

To evaluate whether AI models can replicate humans' natural ability to assess emotions by accurately categorizing emotional states and gauging their intensity, we assess classification accuracy and confidence calibration metrics, including Expected Calibration Error (ECE), Maximum Calibration Error (MCE), and Adaptive ECE (AECE). These metrics align predicted probabilities with emotional intensity, reflecting the system's ability to interpret nuanced affective states. We compare a Rank-O-ToM with existing FER approaches on benchmarks, RAF-DB~\cite{li2017reliable}, FERPlus~\cite{barsoum2016training}, and AffectNet~\cite{mollahosseini2017affectnet} (details in Appendix).

Table~\ref{tab1} shows that Rank-O-ToM outperforms state-of-the-art FER methods on RAF-DB and FERPlus, with comparable results on AffectNet, demonstrating its ability to capture emotional categories and intensities. Additionally, the superior calibration performance underscores the effectiveness of our ranking-based loss in interpreting emotional granularity.

\begin{figure}[t]
\vspace*{-5mm}
  \centering
  \includegraphics[width=0.95\columnwidth]{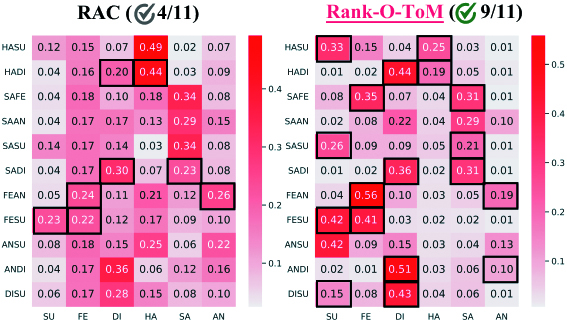}
  \vspace*{-3mm}
  \caption{Confidence heatmaps for RAF-DB compound set: Basic expressions (x-axis) and compound expressions (y-axis) with bold squares marking correct Top-2 matches.}
  \vspace*{-3.5mm}
  \label{fig:heatmap}
\end{figure}

Figure~\ref{fig:CAM} demonstrates how our model interprets emotional expressions using class activation maps (CAMs)~\cite{zhou2016learning} and confidence scores for original and synthetic samples. The CAMs reveal that synthetic samples activate a broader range of facial regions, showcasing a more extensive focus on diverse affective features in Rank-O-ToM. Moreover, the confidence scores indicate that our model is susceptible to complex affective cues, capturing blended and subtle emotions. Figure~\ref{fig:heatmap} illustrates the evaluation of compound emotions (e.g., happily surprised)~\cite{li2017reliable}, assessing whether Top-2 confidence scores match their components (e.g., happiness and surprised). Unlike the existing method, our model effectively aligns confidence scores with constituent emotions, reflecting a nuanced understanding of emotional complexity (details in Appendix).

\vspace*{-1.5mm}
\section{Conclusion} 
Rank-O-ToM advances FER by capturing emotional granularity through synthetic blending and ordinal ranking, aligning AI predictions with human-like reasoning about affective states. Demonstrating superior accuracy and calibration across datasets, it bridges the gap between AI and human cognitive capabilities, enabling nuanced emotional understanding critical for affective ToM. 

\section*{Acknowledgements}
This research was supported by the Institute for Information \& Communications Technology Planning \& Evaluation (IITP) grant funded by the Korea government (MSIT) (No. 2021-0-01341, Artificial Intelligence Graduate School Program, Chung-Ang University).

\bibliography{aaai25}

\appendix

\section{Appendix}

% This appendix provides explanations for additional experiments and visualizations. In \ref{sec:ablations_combine}, we introduce an ablation study on the different schemes for generating synthetic samples. \ref{sec:balanced_confidence} describes visualization and analysis of balanced confidence on basic expression of RAF-DB.
% \ref{sec:appendix_exp} comprises implementation details of the whole experiment and an experiment with a compound set of RAF-DB.
% \ref{sec:additional_CAM} presents additional qualitative results.

This appendix provides additional explanations, analytical experiments, visualizations, and implementation details for the proposed Rank-O-ToM framework. Section~\ref{Details on Synthetic Sample Generation} details the process of integrating the FR dataset and generating synthetic samples. Section~\ref{sec:appendix_exp} includes implementation details for all experiments and an additional experiment using the compound subset of RAF-DB. Finally, Section~\ref{Additional Analysis on Rank-O-ToM} presents further analytical experiments and visualizations to deepen the understanding of Rank-O-ToM's performance and capabilities.

\setcounter{secnumdepth}{2}

\section{Details on Synthetic Sample Generation}
\label{Details on Synthetic Sample Generation}

To enhance the model's ability to interpret complex affective states—a core aspect of affective ToM—we detail the integration of an unlabeled FR dataset, which lacks emotion category labels, and the generation of synthetic samples with varied emotional expressions. These methods address the limitations of relying solely on FER datasets, which may fail to capture the full spectrum of affectivity, leading to overly simplistic comparisons and suboptimal calibration~\cite{wang2023calibration}.

\noindent \textbf{Integrating FR Dataset with Pseudo-labeling.}
Human affective understanding involves interpreting subtle cues and varying emotional intensities across contexts. To emulate this, we integrate an unlabeled FR dataset into training using a confidence-based dynamic thresholding mechanism to assign pseudo-labels. This approach adapts thresholds for each class based on the model's confidence during training, addressing FER data imbalance and enhancing diversity. By incorporating pseudo-labeled FR data, the model learns to represent nuanced and realistic emotional expressions, aligning its learning process with human-like affective reasoning.

First, consider the labeled FER dataset $\mathcal{D}_{\mathrm{fer}}=\{(x_{i}^{\mathrm{fer}},y_{i}^{\mathrm{fer}})\}_{i=1}^{n}$. For each sample, we obtain the predicted label $\hat{y}_i^{\mathrm{fer}}$ for the $i$-th sample at epoch $t$. To determine class-specific confidence thresholds, we define the set of correctly predicted samples for each class $c$ as $\mathcal{D}^{\ cs}_{\mathrm{fer}}=\{(x^{\mathrm{fer}}, y^{\mathrm{fer}})|\hat{y}^{\mathrm{fer}}= y^{\mathrm{fer}}= c\}$. For each sample in this set, we calculate the confidence score $\tilde{p}^{\mathrm{fer}} = \max_{c} p_c(y^{\mathrm{fer}} | x^{\mathrm{fer}})$. For $(x^{\mathrm{fer}}, y^{\mathrm{fer}}) \in \mathcal{D}^{\ cs}_{\mathrm{fer}}$, we calculate the confidence $\tilde{p}^{\mathrm{fer}} := \underset{c}{\mathrm{max}}\ {p_{c}(y^{\mathrm{fer}}|x^{\mathrm{fer}})}$. For each epoch $t$ and class $c$, the confidence threshold $\mathcal{T}_{c}^{t}$ is computed as follows:
\begin{align}
    \mathcal{T}_{c}^{t}&=\frac{\beta}{1+e^{-t}} \cdot \frac{1}{|\mathcal{D}^{\ cs}_{\mathrm{fer}}|} \sum_{i=1}^{|\mathcal{D}^{\ cs}_{\mathrm{fer}}|}{\tilde{p}_i^{fer}} \notag\\
    &=\frac{\beta}{1+e^{-t}} \cdot \frac{1}{|\mathcal{D}^{\ cs}_{\mathrm{fer}}|}\sum_{i=1}^{|\mathcal{D}^{\ cs}_{\mathrm{fer}}|}{\underset{c}{\mathrm{max}}\ p_{c}(y_{i}^{\mathrm{fer}}|x_{i}^{\mathrm{fer}})}
    \notag\\& \qquad  ,(x_{i}^{\mathrm{fer}}, y_{i}^{\mathrm{fer}}) \in \mathcal{D}^{\ cs}_{\mathrm{fer}}
    \label{eq:4}
\end{align}

The hyperparameter $\beta \in (0, 1)$ moderates confidence levels. The term $\frac{1}{1 + e^{-t}}$ ensures that the threshold adapts dynamically over training epochs, reflecting the model's evolving understanding—analogous to how human perception becomes more refined with experience.

During training, FER models exhibit varied patterns across classes due to data imbalance~\cite{li2022towards}. Initially, there's a narrow confidence gap between classes with abundant and scarce samples. As training advances, this distinction becomes more pronounced, encompassing both clear and ambiguous emotional expressions. The dynamic threshold $\mathcal{T}_c^t$ adjusts to these shifts, enabling the model to calibrate its confidence in a manner akin to human affective judgment.

Subsequently, for the unlabeled FR dataset $\mathcal{D}_{\mathrm{fr}}=\{x_{i}^{\mathrm{fr}}\}_{i=1}^{m}$, we use distinct weak augmentations to obtain two transformed samples, $x_{\mathtt{a}}^{\mathrm{fr}}=\mathtt{Aug_a}(x_i^{fr})$ and $x_{\mathtt{b}}^{\mathrm{fr}}=\mathtt{Aug_b}(x_i^{fr})$. Then, through the model $\mathcal{F}(\cdot)$, we determine the probability distributions $p_{\mathtt{a}}^{\mathrm{fr}}$ and $p_{\mathtt{b}}^{\mathrm{fr}}$ for facial expressions' classes for the two transformed samples $x_{\mathtt{a}}^{\mathrm{fr}}$ and $x_{\mathtt{b}}^{\mathrm{fr}}$. To emulate the human ability to integrate multiple subtle cues when interpreting emotions, we perform class-wise interpolation to obtain the aggregated probability distribution $\hat{p}$ for each sample:
\begin{align}
    \hat{p}_c=\lambda_c\cdot p_{\mathtt{a}}^{\mathrm{fr}}(c|x_{\mathtt{a}}^{\mathrm{fr}}) + (1-\lambda_c)\cdot p_{\mathtt{b}}^{\mathrm{fr}}(c|x_{\mathtt{b}}^{\mathrm{fr}})
    \label{eq:5}
\end{align}

Here, $c$ represents the class in FER, and $\lambda_c$ denotes the interpolation ratio for class $c$. We assign $\hat{y}^{\mathrm{fr}}$ based on the class with the highest confidence among those exhibiting confidence higher than the threshold $\mathcal{T}_c^t$, as follows:
\begin{align}
\hat{y}_i^{\mathrm{fr}}=\mathrm{argmax}_{c}(\mathds{1}(\hat{p}_c>\mathcal{T}_c^{t})\cdot\hat{p}_c)
    \label{eq:6}
\end{align}

where $\mathds{1}(\cdot)$ indicates whether the probability score belonging to a specific class $c$ is larger than the threshold $\mathcal{T}_c^t$. This method enables the integration of the unlabeled FR dataset during training, increasing the diversity of synthetic samples and enhancing the efficiency of ranking relationships. It allows the model to learn from a wider array of facial expressions, capturing subtle variations and complexities—mirroring the human capacity for nuanced affective understanding central to affective ToM.

\begin{figure*}[t!]
  \centering
  \includegraphics[width=0.95\textwidth]{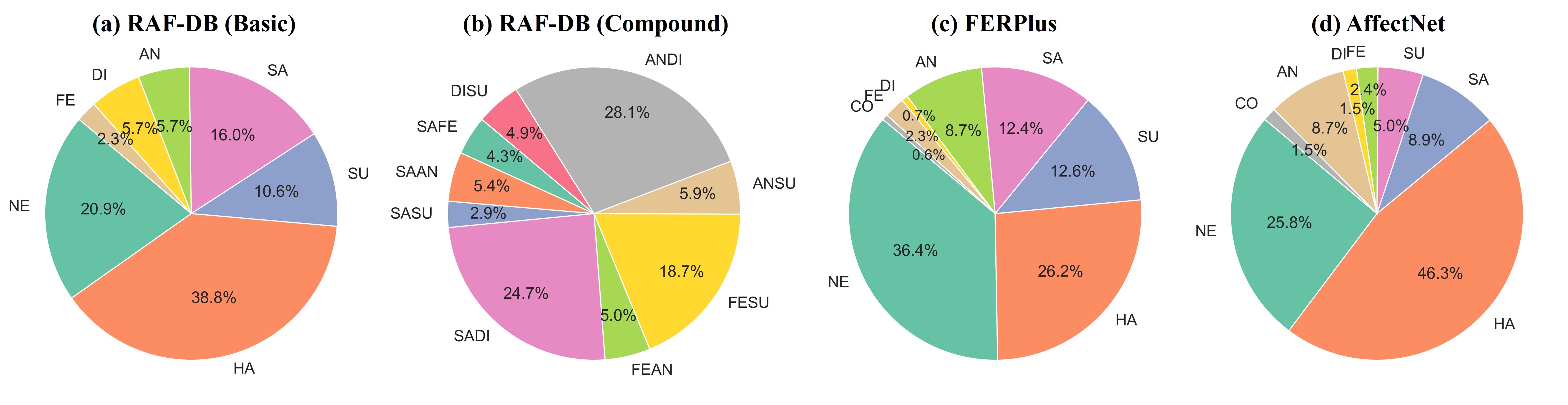}
  \caption{{\textbf{Class proportions of FER datasets} The pie charts display results for (a) RAF-DB (Basic), (b) RAF-DB (Compound), (c) FERPlus, (d) AffectNet.}}
  \label{fig:sup_class_ratio}
\end{figure*}

\begin{table*}[t!]
\centering
\resizebox{\linewidth}{!}{%
% \large{
\setlength{\tabcolsep}{6pt}
\renewcommand{\arraystretch}{1.1}
\begin{tabular}{ll|c|c|l}
\Xhline{4\arrayrulewidth}
\multicolumn{2}{c|}{Datasets}                                                     & {\begin{tabular}[c]{@{}c@{}}\# Annotated images\\ (train / val / test)\end{tabular}}      & \# Classes & Emotional labels                                                                                                                                                                                                                                                        \\ \hline \hline
\multicolumn{1}{l|}{\textbf{RAF-DB}~\shortcite{li2017reliable}}  & Basic                            & {\begin{tabular}[c]{@{}c@{}}15,339\\ (12,271 / 3,068 / -)\end{tabular}}  & 7          & neutral, happiness, surprise, sadness, anger, disgust, fear                                                                                                                                                                                                            \\ \cline{2-5}
\multicolumn{1}{l|}{} & Compound & {\begin{tabular}[c]{@{}c@{}}3,954\\ (3,162 / 792 / -)\end{tabular}} & 11         & \begin{tabular}[c]{@{}l@{}}happily surprised, happily disgusted, sadly fearful, sadly angry,\\ sadly surprised, sadly disgusted, fearfully angry, fearfully surprised,\\ angrily surprised, angrily disgusted, disgustedly surprised\end{tabular} \\ \hline
\multicolumn{2}{l|}{\textbf{FERPlus}~\shortcite{barsoum2016training}}                                                     & {\begin{tabular}[c]{@{}c@{}}35,887\\ (28,558 / 3,579 / 3,573)\end{tabular} } & 8          & neutral, happiness, surprise, sadness, anger, disgust, fear, contempt \\ \hline
\multicolumn{2}{l|}{\textbf{AffectNet}~\shortcite{mollahosseini2017affectnet}}                                                   & {\begin{tabular}[c]{@{}c@{}}291,650\\ {(287,651 / 3,999 / -)}\end{tabular}}        & 8          & neutral, happiness, surprise, sadness, anger, disgust, fear, contempt \\ \Xhline{4\arrayrulewidth}
\end{tabular}
}
% }
\caption{\textbf{Overview of FER datasets} This presents the datasets used to evaluate FER performance. It includes the number of annotated images, the number of classes, and the emotional labels used in experiments. The datasets covered are RAF-DB (both Basic and Compound), FERPlus, and AffectNet.}
\label{tab:FER_dataset}
\end{table*}

\noindent \textbf{Synthesizing Samples with Diverse Emotional Expressions.} To enhance the model's ability to interpret complex affective states, we generate synthetic samples that capture a broader range of emotional expressions by blending facial images with different labels. This emulates the human ability to perceive blended or intermediate emotions, which are often absent in FER datasets.

For example, combining a sample labeled ``sadness" with one labeled ``anger" creates an image representing an intermediate emotion like ``frustration." This aligns with the human capacity to interpret compound emotions, a key aspect of affective ToM. We achieve this using an adapted CutMix augmentation method~\cite{yun2019cutmix}, which preserves the semantic integrity of facial expressions. CutMix is an augmentation technique that generates synthetic samples by cutting and pasting regions from two training images. Given two samples, $(x_i, y_i)$ and $(x_j, y_j)$, CutMix creates a new sample $(\tilde{x}, \tilde{y})$ as follows:

\begin{align} \tilde{x} = M \odot x_i + (1 - M) \odot x_j, \quad \tilde{y} = \lambda y_i + (1 - \lambda) y_j, \label{eq:cutmix} \end{align}

where $M$ is a binary mask indicating the region to be replaced, $\odot$ denotes element-wise multiplication, and $\lambda \sim Beta(1, 1)$ controls the combination ratio, ensuring a balanced mix of the two images. The mask $M$ is determinded by a randomly generated bounding box $B = (b_x, b_y, b_w, b_h)$, where: 

\begin{align} b_x &\sim \text{Uniform}(0, W), \quad b_w = W \sqrt{1 - \lambda}, \\ \ b_y &\sim \text{Uniform}(0, H), \quad b_h = H \sqrt{1 - \lambda}. \label{eq:cutmix_bbox} \end{align}

In standard CutMix, the position and size of the bounding box $B$ are randomly determined, which may disrupt the semantic structure of facial expressions critical for emotion recognition. 

To preserve the semantic regions associated with different emotional expressions—such as the eyes and mouth—we adapt the CutMix method by fixing the bounding box to horizontally split the image. Specifically, we set the top-left coordinates to $b_x=0$ and $b_y=0$, and define the width and height as $b_w=W$ and $b_h = 1/2H$, respectively. This effectively divides the face into upper and lower halves, each containing distinct emotional cues. Since both facial regions contribute equally to the combined image, we fix the combination ratio $\lambda=0.5$. This ensures that the synthetic sample $\tilde{x}$ integrates the upper half from one image and the lower half from another, maintaining the balance of emotional expressions. 

By horizontally bisecting and combining facial images in this manner, we preserve the critical semantic information necessary for affective understanding. This approach aligns with the human ability to integrate facial cues from different areas of the face—a key aspect of affective ToM.

\section{Experimental Details}
\label{sec:appendix_exp}

\subsection{Datasets}
We train Rank-O-ToM on the FER benchmark datasets, including RAF-DB~\cite{li2017reliable}, FERPlus~\cite{barsoum2016training}, and AffectNet~\cite{mollahosseini2017affectnet}. Additionally, we incorporate the LFW face recognition dataset~\cite{huang2008labeled}, which does not include emotion class labels, into our training process. Overview of each datasets is shown in Table~\ref{tab:FER_dataset}, and the proportions of each dataset are shown in Figure~\ref{fig:sup_class_ratio}.

\noindent \textbf{RAF-DB}~\cite{li2017reliable} is a facial expression dataset comprising 29,672 individual facial images, including basic or compound expressions. In this work, we utilize facial images with 7 expressions (i.e., ``surprise'', ``fear'', ``disgust'', ``happiness'', ``sadness'', ``anger'', ``neutral''), including 12,271 images as training data and 3,068 images as test data. 
In addition, we employ the \textit{compound set} for validation in Figure~\ref{fig:heatmap} to evaluate our framework's performance on more complex and nuanced emotional expressions. The compound dataset comprises 11 distinct classes, each representing a combination of basic emotions (i.e., ``happily surprised'', ``happily disgusted'', ``sadly fearful'', ``sadly angry'', ``sadly surprised'', ``sadly disgusted'', ``fearfully angry'', ``fearfully surprised'', ``angrily surprised'', ``angrily disgusted'', and ``disgustedly surprised''). 
Images are aligned and cropped using three landmarks, then resized to $224 \times 224$ pixels.

\noindent \textbf{FERPlus}~\cite{barsoum2016training} is an extended version of the original FER dataset, designed to improve label accuracy and reliability through enhanced annotations provided by 10 annotators via crowdsourcing. It includes 35,887 facial images labeled with 8 basic expressions: ``neutral'', ``happiness'', ``surprise'', ``sadness'', ``anger'', ``disgust'', ``fear'', and ``contempt''. Additionally, FERPlus utilizes a multi-label annotation, allowing images to be associated with multiple emotions, reflecting the complexity of human facial expressions. Images are aligned and cropped using key facial landmarks, then resized to $224 \times 224$ pixels for consistency with our processing pipeline.

\noindent \textbf{AffectNet}~\cite{mollahosseini2017affectnet} is one of the most comprehensive and sizable facial expression datasets, containing 287,651 training images and 3,999 test images manually labeled into eight classes. While AffectNet includes a wide range of annotations, in this work, we utilize only the images labeled with eight basic emotion categories: ``neutral'', ``happiness'', ``surprise'', ``sadness'', ``anger'', ``disgust'', ``fear'', and ``contempt''. This selection excludes images with labels such as ``none'', ``uncertain'', and ``non-face'', which do not correspond to any specific emotion.

\noindent \textbf{LFW}~\cite{huang2008labeled} is a popular FR dataset that has not been labeled for emotion categories. LFW comprises 13,000 facial images, representing over 5,000 identities. In our study, we pseudo-label the LFW dataset into seven emotion classes, the same as the RAF-DB dataset.
We utilize the MTCNN~\cite{zhang2016joint} alignment method to detect and align faces within the LFW dataset, resizing them to $224 \times 224$ pixels.

\subsection{Evaluation Metrics}

When evaluating AI agents in the context of affective ToM, it is crucial to assess not only their accuracy but also how well their confidence—derived from softmax scores—aligns with the true likelihood of their predictions. This alignment, quantified through calibration metrics, ensures that the model's confidence levels are appropriate for reliable human-AI interactions, particularly when interpreting nuanced emotional expressions.

To assess calibration, predictions are grouped into $M$ bins of equal size intervals. For each bin $B_m$ containing samples whose confidence scores fall within bin $m$, we define accuracy and confidence as follows:
\begin{align}
    \mathrm{acc}(B_{m})&=\frac{1}{|B_m|}\sum_{i\in B_{m}}{\mathds{1}(\hat{y}_{i}=y_i)} \\
    \mathrm{conf}(B_{m})&=\frac{1}{|B_m|}\sum_{i\in B_{m}}{\tilde{p}_i}
    \label{eq:10}
\end{align}
where $\tilde{p}_i$ represents the confidence of sample $i$. 

\noindent \textbf{Expected Calibration Error (ECE)} measures the average discrepancy between a model's accuracy and confidence across bins:
\begin{align}
    \mathrm{ECE} = \sum_{m=1}^{M}\frac{|B_m|}{n}{\mid \mathrm{acc(B_{m})}-\mathrm{conf}(B_{m})\mid}
    \label{eq:11}
\end{align}

\noindent \textbf{Adaptive Expected Calibration Error (AECE)} extends ECE by dynamically adjusting the binning process to better match the distribution of predicted probabilities, offering a finer-grained evaluation of calibration.

% \begin{equation*}
% \operatorname{AECE} = \sum_{m=1}^{M}\frac{|B_m|}{n}\left|\operatorname{acc(B_m)}-\operatorname{conf(B_m)}\right|.
% \end{equation*}

\noindent \textbf{Maximum Calibration Error (MCE)} highlights the largest miscalibration by capturing the maximum discrepancy between accuracy and confidence across bins:
\begin{align}
    \mathrm{MCE} = \operatorname{max}_{m\in(1,...M)}{\mid \mathrm{acc(B_{m})}-\mathrm{conf}(B_{m})\mid}
    \label{eq:12}
\end{align}

These metrics enable the evaluation of how well a model's confidence aligns with its true performance, ensuring reliable and interpretable predictions. Such alignment is crucial for interpreting complex human emotions, as it allows the model to emulate human-like reasoning and maintain appropriate confidence levels in affective ToM tasks.

\subsection{Implementation Details}

% \noindent \textbf{Implementation details} 
We utilize the Adam optimizer~\cite{2015-kingma} with a learning rate of $5 \times 10^{-4}$. Training proceeds with a mini-batch size of 64 over 60 epochs. The initial threshold for class-wise dynamic pseudo-labeling starts at 0.95. Weak augmentations apply, including RandomCrop and RandomHorizontalFlip. Hyperparameters stand at $\lambda_c = \frac{1}{2}$ and $\beta = 0.97$.  We evaluate the performances on calibration metrics, setting all bins to 15.
Ranking loss uses a margin of $\delta \in [0.1,0.3]$ and a balancing scalar $\beta=1$, which balances the focal loss, which is employed to classify emotion categories as the default values unless otherwise specified. The hyperparameter $\gamma=2$ and $\alpha=0.25$ are used in focal loss~\cite{lin2017focal}.

\noindent \textbf{Implementation details on an experiment with a compound set of RAF-DB}
We elaborate on the details of the experiment regarding compound emotions in Rank-O-ToM. We utilize a compound RAF-DB dataset composed of compound expressions. The compound RAF-DB consists of data annotated with compound expressions where two basic expressions (e.g., ``sad'' and ``angry'') are combined to form a compound expression (e.g., ``sadly angry''). A well-trained FER framework should demonstrate high confidence levels in the basic expressions that comprise the compound expression when presented with samples of compound expressions, provided it has learned a diverse spectrum of expressions. For instance, with a sadly angry'' sample, a well-calibrated FER framework trained with basic emotions should output high probabilities for ``sad'' and ``angry.''

\begin{figure*}[t!]
  \centering
  \includegraphics[width=0.95\textwidth]{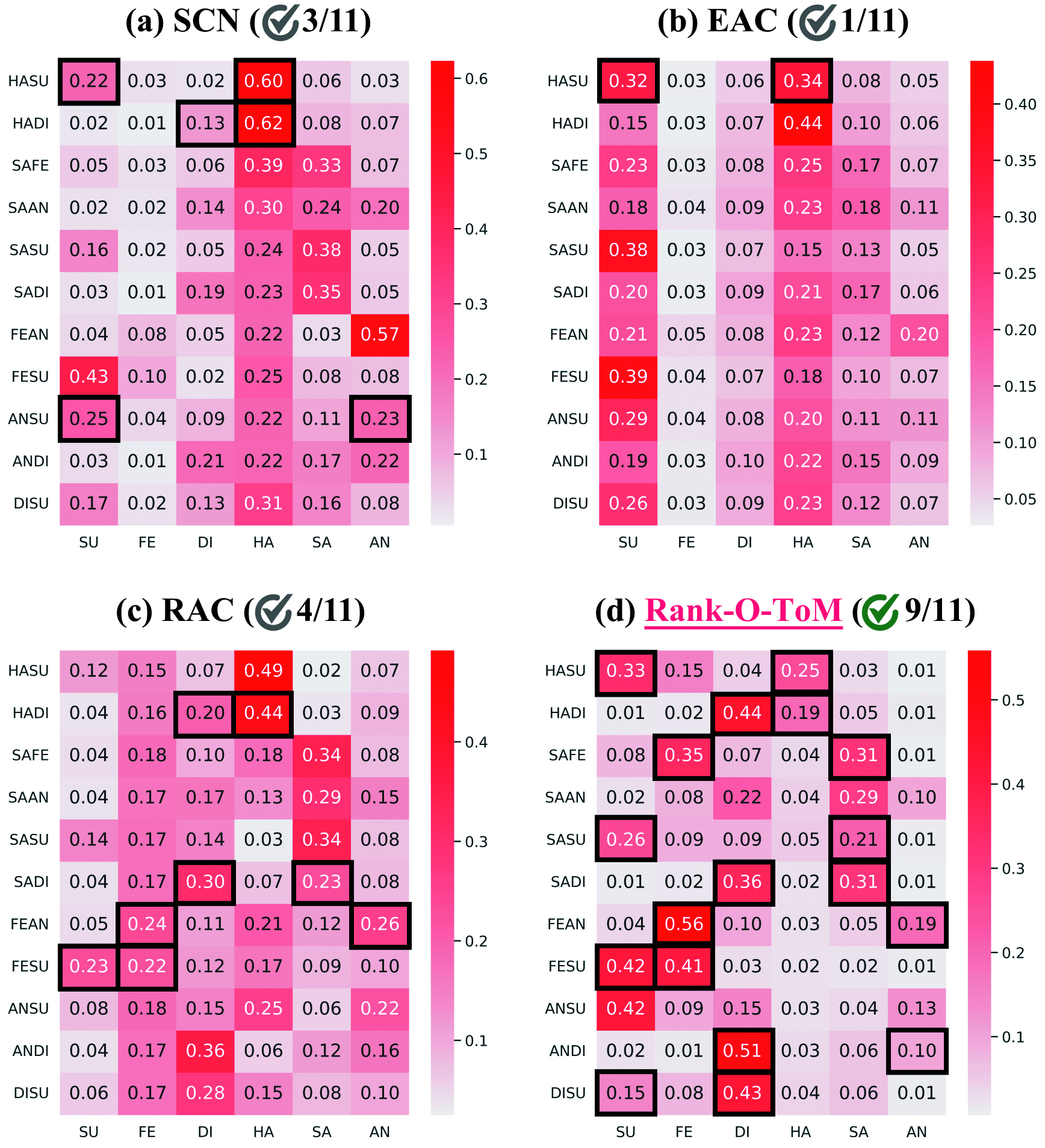}
  \caption{\textbf{Additional heatmaps for RAF-DB compound set}. Basic expressions (x-axis) and compound expressions (y-axis) are depicted, with bold squares marking correct Top-2 matches. SCN and EAC models are presented here as supplementary to the main text, which focuses on RAC and Rank-O-ToM.}
  \label{fig:compound_confidence_heatmap_all}
\end{figure*}

The heatmaps in Figure~\ref{fig:heatmap} and \ref{fig:compound_confidence_heatmap_all} depict the average confidence for samples containing compound expressions, as evaluated by both FER methods and Rank-O-ToM on the RAF-DB dataset. Additionally, we evaluate whether the Top-2 confidence scores inferred by the model match the basic emotions constituting the compound expression. Experimental results indicate that Rank-O-ToM activates the constituent expressions in a balanced manner compared to other approaches.

% \section{Analysis of balanced confidence of FER methods}
\section{Additional qualitative examples on CAM}
\label{Additional Analysis on Rank-O-ToM}

\begin{sidewaysfigure*}
\centering
  \includegraphics[width=0.9\textwidth]{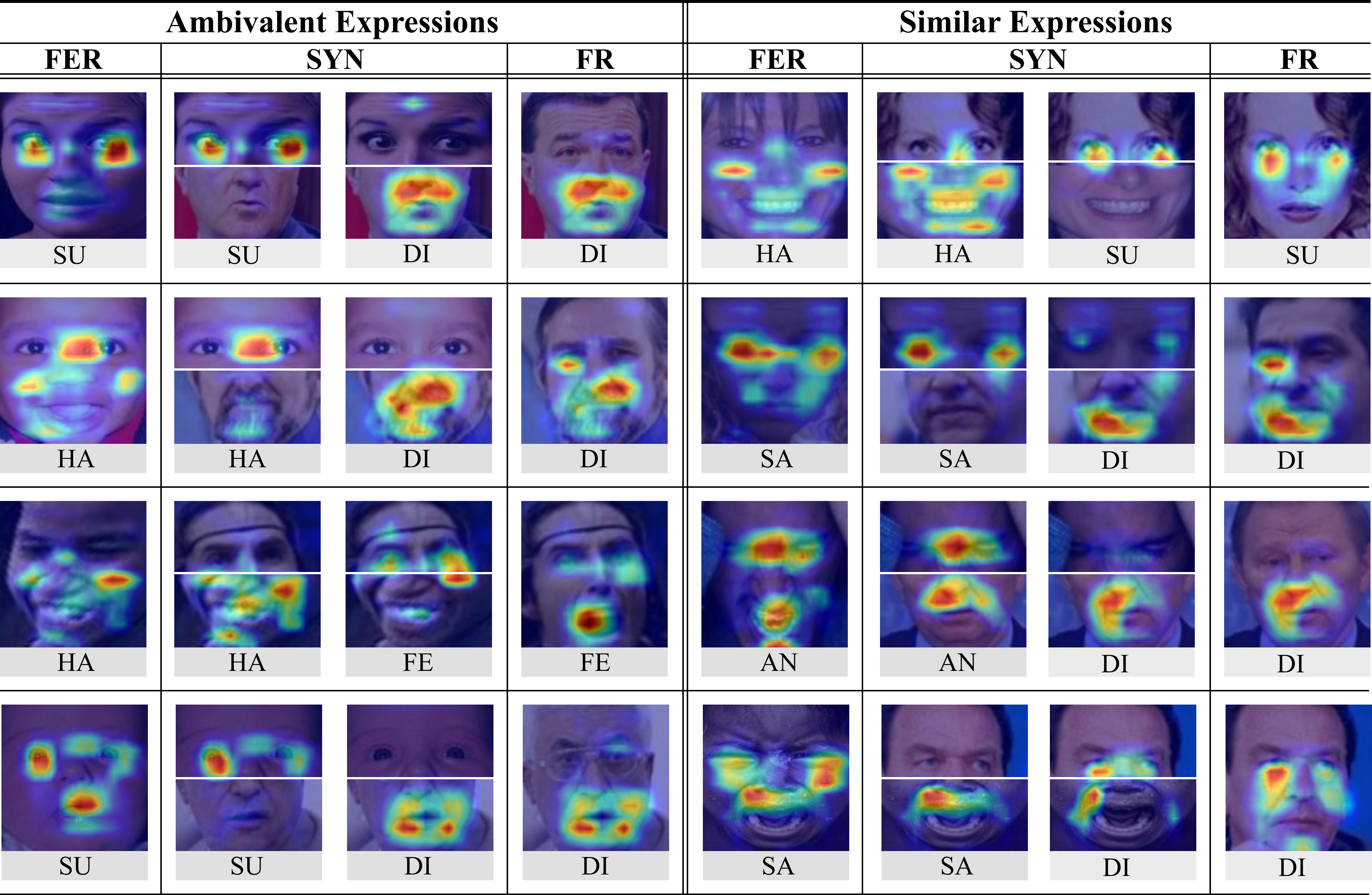}
    \caption{\textbf{Qualitative results of CAM on original samples and synthetic samples generated by Rank-O-ToM}}
  \label{fig:sup_CAM}
\end{sidewaysfigure*}

In Figure~\ref{fig:sup_CAM}, we present additional qualitative results on the CAMs produced by our framework for two original samples—from the FER and FR datasets—and the synthetic sample (SYN) generated by combining them. These visualizations illustrate various combinations of emotions, demonstrating how our method captures nuanced affective states essential for human-like emotion recognition.

\end{document}